\documentclass[12pt,pra,aps,amssymb,amsfonts,amsmath,tightenlines]{revtex4}

\usepackage[dvips]{graphicx}
\usepackage{dsfont, color,txfonts} 

\begin{document}

\title{Noise, fake news, and tenacious Bayesians}
\author{Dorje C. Brody}%

\affiliation{
Department of Mathematics, University of Surrey, Guildford GU2 7XH, UK 
}

\date{\today}

\begin{abstract}
A modelling framework, based on the theory of signal processing, for characterising the dynamics of systems driven by the unravelling of information is outlined, and is applied to describe the process of decision making. The model input of this approach is the specification of the flow of information. This enables the representation of (i) reliable information, (ii) noise, and (iii) disinformation, in a unified framework. Because the approach is designed to characterise the dynamics of the behaviour of people, it is possible to quantify the impact of information control, including those resulting from the dissemination of disinformation. It is shown that if a decision maker assigns an exceptionally high weight on one of the alternative realities, then under the Bayesian logic their perception hardly changes in time even if evidences presented indicate that this alternative corresponds to a false reality. Thus confirmation bias need not be incompatible with Bayesian updating. By observing the role played by noise in other areas of natural sciences, where noise is used to excite the system away from false attractors, a new approach to tackle the dark forces of fake news is proposed.   
\end{abstract}

\maketitle


\section{Introduction}

The term `fake news' traditionally is understood to mean false newspaper 
stories that have been fabricated to enhance the sales of the paper. While 
unethical, in most cases they are not likely to create long-lasting serious 
damages to society. However, since the 2016 US presidential election 
and the 2016 `Brexit' referendum in the UK on the membership of the 
European Union, this phrase has been quoted more frequently, with the 
understanding that it 
refers to deliberate disseminations of false information with an intent to 
manipulate the public for political or other purposes. The concept of fake 
news in the latter sense, of course, has been around for perhaps as long as 
some three 
thousand years, and historically it has often been implemented in the context 
of conflicts between nations or perhaps even between corporations. Hence 
there is nothing new in itself about fake news, except that the rapid 
development of the 
Internet over the past two decades has facilitated its application in major 
democratic processes in a way that has not been seen before, and this has 
not only attracted attention of legislators (Collins \textit{et al}. 2019, 
Gallo \& Cho 2021) but 
also generated interests in academic studies of the phenomenon, its 
implications and prevention 
(Allcott \& Gentzkow 2017, Shu \textit{et al}. 2017, Bastos \& Mercea 2018, 
Sample \textit{et al}. 2019, Boovet \& Makse 2019, Scheufele \& Krause 2019 
Grinberg \textit{et al}. 2019, Rajabi \textit{et al}. 2019, Connor Desai \textit{et al}. 
2020, Roozenbeek \textit{et al}. 2020, to name a few). The 
purpose of the present paper is to contribute towards this endeavour by 
applying techniques of communication theory to develop a general framework 
for characterising the dynamical behaviours of systems (for 
example, a group of people) driven by the flow 
of information, irrespective of whether the information is true or false. 

The idea that social science, more generally, can only be properly understood 
by means of communication theory, for, communication is the building block 
of any community and hence society, was advocated by Wiener long ago 
(Wiener 1954), although its practical implementation has only been developed 
over the past twenty years in the context of financial applications (Brody 
\textit{et al}. 2022). When it comes to the study of the impact of information 
revelation, whether the information is reliable or not, in particular, the 
techniques of communication theory become especially effective. 
This follows from the observation that the change in the 
behaviour of a decision maker that we intend to model results directly from 
communicating information. 
Based on this observation, a systematic investigation of the effects of fake 
news in the context of electoral competition and referendum, from the 
viewpoint of information transmission, was initiated in Brody \& Meier (2022; 
first appeared in 2018) 
and developed further in Brody (2019). The concepts underlying these works 
are elaborated here in greater detail, with the view towards 
developing measures to counter the negative impacts of 
deliberate or malicious disinformation that misguide the public. 

In more specific terms, to study the impact of disinformation, it is 
indispensable that information, noise such as rumours and speculations, 
disinformation, the rate of information revelation, and so on, are all 
represented by quantities that take numerical values. 
Otherwise, scientifically meaningful analysis, such as determining the 
likelihood of certain events to take place, cannot be applied.
In probability theory, this idea is represented by the 
concept of random variables that assign numerical values to outcomes of 
chance. To study the 
impact of disinformation, or more generally to study the dynamics of a system 
governed by information revelation, therefore, the information-providing random 
time 
series (which may or may not contain disinformation) will be modelled. Given 
this `information process' it is 
then possible to apply powerful and well established techniques of 
communication theory to study virtually all dynamical properties of the 
system, including the statistics of future events. In fact, as shown below, 
the method is sufficiently versatile to the extent that it allows for the 
numerical simulation of an event that occurs with zero probability -- a simulation 
of what one might call an alternative fact. The fundamental idea underpinning 
the present approach is that if a decision maker were 
to follow Bayesian logic (Bayes 1763) for assessing uncertain events, then 
the statistics of their 
behaviour can be predicted from a simple mathematical deduction, provided that 
the flow of information is specified. This motivates us to model 
the information flow as the starting point so as to \textit{derive} the dynamical 
behaviours of people driven by information revelation. This is in contrast to more 
traditional approaches in mathematical modelling whereby one attempts to 
model the behaviour itself from the outset. The latter approach is problematic 
in the context of information-driven systems under noisy environments, for, 
the dependence of the output (behavioural dynamics) on the input (information 
revelation) is often highly nonlinear. 

With this in mind the present paper explains how the flow of information can be 
modelled, and how the unravelling of information under noisy environments 
affects a decision maker's perception. Then it is shown how the model can 
be applied to determine the dynamics of an electoral competition, and, in 
particular, how a deliberate dissemination of disinformation might affect the 
outcome of a future election. The two fundamental ways in which the 
information can be manipulated will be discussed. The paper then introduces 
the concept, to be referred to as the tenacious Bayesian, that explains how 
people behave in a seemingly irrational manner if they excessively 
overweight their beliefs on 
a false reality, even though they are following the rational 
Bayesian logic. This shows that an element of confirmation 
bias can be explained within the Bayesian framework, contrary to what is 
often asserted in the literature. 
Finally, the paper proposes a new approach to counter the 
impact of disinformation, by focusing on the role played by noise, and by 
borrowing 
ideas from statistical physics of controlling a system that entails many 
internal conflicts or frustrated configurations. 
Specifically, it is common for a complex system to be 
trapped in a locally stable configuration that is globally suboptimal, 
because the system has to enter into a highly unstable configuration 
before it can reach an even more stable one. However, by increasing 
the noise level the system is excited and becomes unstable, thence 
by slowly reducing the noise level the system has a chance of 
reaching an even more stable configuration.

\section{Decision dynamics from information processing}

Decision making arises when one is not 100\% certain about the `right' 
choice, due to insufficient information. The current knowledge relevant to 
decision making then reflects the \textit{prior} uncertainty. If additional partial 
information about the quantity of interest arrives, then this \textit{prior} is 
updated to a \textit{posterior} uncertainty. To see how this 
transformation works it suffices to 
consider a simple example of a binary decision -- a decision between two 
alternatives labelled by $0$ and $1$ -- under uncertainty. Suppose 
that we let $X$ be the random variable representing a binary decision so that 
$X$ takes the value $0$ with probability $p$ and $X$ equals $1$ with 
probability $1-p$, where the probabilities reflect the degree of uncertainty. 
In the context of an electoral competition, one can think of a two-candidate 
scenario whereby $X=0$ corresponds to candidate $A$ and $X=1$ 
corresponds to candidate $B$. Then the probabilities $(p,1-p)$ reflect the 
\textit{a priori} view of a given decision maker -- a voter for example. In 
particular, if $p>0.5$, then candidate $A$ is currently preferred over 
candidate $B$. 

With this setup, the decision maker receives additional noisy information about 
the `correct' value of $X$. For example, one might read an 
article that conveys the information that voting for candidate $A$ is likely to be 
the correct decision. The idea then is to translate this information into a 
numerical value so as to be able to understand and 
characterise how the view of 
the decision maker, represented by the probabilities $(p,1-p)$, is affected by 
acquiring further information. To model this mathematically, let $\epsilon$ 
denote the random variable representing noise, which is assumed statistically 
independent of $X$. The origin of noise may be a genuine 
mistake, or a plain speculation, on the part of the author of the article, or 
perhaps a 
simple misinterpretation of the article on the part of the decision maker. 
The idea thus is to regard the unknown 
quantity of interest, the value of 
$X$ in this case, as a signal to be inferred, which is obscured by noise. Hence 
the receiving of noisy information amounts to observing the value of 
\[ 
\xi = X + \epsilon . 
\] 
Because there are two unknowns, $X$ and $\epsilon$, and one known, the 
value of $\xi$, it is not possible to determine the value of $X$, 
which reveals the correct choice of action, from this 
information. Nevertheless, the knowledge of the value of $\xi$ will allow the 
decision maker to reduce the uncertainty about $X$. As an example, 
suppose that there is a wide range 
of rumours and speculations about the value of $X$. If there are many 
such contributions to the noise, then the law of large numbers implies that it 
is reasonable to assume $\epsilon$ be normally distributed, say, with mean 
zero and some standard deviation $\nu$. Of course, the 
nature of noise may not be of Gaussian type, and likewise the signal and 
noise decomposition in general need not be additive. One of the advantages 
of the present approach is that depending on the context, it is possible to 
choose the structure of the information-baring random variable $\xi$, and 
proceed to analyse its consequences (see, e.g., Brody \textit{et al}. 2013). 
However, for illustrative purposes here we shall proceed with the additive 
Gaussian noise model.  

Suppose that the value of $\nu$ is relatively small, say, $\nu=0.2$. This means 
that the distribution of $\epsilon$ is narrowly peaked at $\epsilon=0$. Suppose 
further that the value of the observation is $\xi=0.73$. In this case there are 
two possibilities: we have either $(X,\epsilon)=(0,0.73)$ or $(X,\epsilon)=
(1,-0.27)$. However, given that the distribution of $\epsilon$ is narrowly peaked 
at $\epsilon=0$, the realisation $\epsilon=0.73$ is significantly less likely 
as compared to the event that $\epsilon=-0.27$. Hence after the observation 
that $\xi=0.73$ the prior probabilities $(p,1-p)$ will be updated to the posterior 
probabilities $(p',1-p')$ in such a way that $p'<p$, whenever $\xi>0.5$. 
The exact 
value of $p'$ will be dependent on the value of $p$, and can be calculated 
using the Bayes formula: 
\[
p'=\frac{p\,\rho(\xi|X=0)}
{p\,\rho(\xi|X=0)+(1-p)\,\rho(\xi|X=1)},
\]
where $\rho(\xi|X=0)$ is the density function of the random variable $\xi$ 
given the event that $X=0$, and similarly for $\rho(\xi|X=1)$. Because 
conditional on the value of $X$ the random variable $\xi$ is normally 
distributed with mean $X$ and variance $\nu^2$, in the present context the 
Bayes formula gives 
\[
p' = \frac{p}{p+(1-p)\exp\left( \frac{1}{\nu^2}(\xi-\frac{1}{2})\right)} . 
\]
Thus, for instance, if the \textit{a priori} probability is 50-50 
so that $p=0.5$ then we find in this example with 
$\nu=0.2$ and $\xi=0.73$ that $p'\approx0.0032$. In other 
words, although the value of $X$ remains unknown, we can be almost  
(99.68\%) certain that the decision corresponding to 
$X=1$ is the correct choice, based on the observation of 
information relevant to decision making. 
The example here is consistent with our intuition, owing to the fact that our 
brains appear 
capable of subconsciously implementing the Bayes formula at an 
intuitive level, when it comes to processing signals under 
noisy environments --- for example, in attempting to catch a ball in the air, 
instead of consciously solving Newton's equations, our brain subconsciously 
processes the visual signal to achieve the task. (Whether a human brain is 
capable of subconsciously implementing Bayesian rules in broader contexts 
might be questionable --- one recent proposal (Sanborn and Chater 2016) 
is that the brain functions instead as a Bayesian sampler.) 
Therefore, once the observation is made, our views will be 
shifted, resulting in actions, such as making a decision. In other words, it is 
the processing of noisy information that results in the dynamics of decision 
makers: new information arrives, positions reassessed, and actions taken. 

The approach taken here to model the dynamics of decision making is based 
on the standard formalism of communication theory (Wiener 1948, Shannon 
\& Weaver 1949). In communication theory, 
the random variable $X$ represents the signal that has been transmitted, 
which is obscured by noise, represented here by the random variable 
$\epsilon$. The task then is to determine the best estimate of $X$ given the 
observation $\xi$. Because the processing of imperfect information is intrinsic 
to any decision making under uncertainty, communication theory is 
highly effective in characterising dynamical 
behaviours of people driven by information 
revelation. Indeed, communication theory has been applied extensively to 
model dynamical behaviours of financial markets, or more precisely the 
dynamics of asset prices, over the past two decades (Brody \textit{et al}. 
2022). In the context of a financial market, asset prices change in 
accordance with transaction decisions. When a market participant makes a 
decision on an investment, their primary concern is the unknown future return 
resulting from that investment. By letting $X$ be the random variable 
representing 
the return of a given investment over a period, whose value is obscured by 
market noise, it is then possible to arrive at a plausible model for the price 
dynamics using the techniques of signal processing in communication theory, 
because the model merely replicates, albeit with some simplifying 
approximations, what actually takes place in real world -- prices change in 
accordance with the flow of information. 

Traditional communication theorists have shied away from applying techniques 
of signal detection to model behavioural dynamics, for, the random variable 
$X$ appearing in the context of decision making is typically not `transmitted' 
by the sender of a communication channel. Instead, it 
represents the quantity of interest that one wishes to infer under uncertainty. 
In a financial context, for instance, 
$X$ may represent the future return over an investment period, whose value is 
not known to anyone, so clearly no one is able to transmit the value of $X$. 
Yet, $X$ certainly exists, whose value will be revealed at the end of that 
investment period. In this case, it is the market as a whole that acts like an 
abstract communication channel. Likewise, situations are similar for many other 
decision makings under uncertainties, but it requires a leap of imagination to 
realise that communication theory provides a unified framework for 
characterising the dynamics of information-driven systems even when there is 
no explicit mechanical device to form a communication channel. 

There is another reason why, in spite of its effectiveness, signal processing 
had not been widely 
applied to modelling behavioural dynamics, and this has to do 
with the meaning of random variables in probability. Take, for instance, the 
case of coin tossing. If the coin is fair, then the outcome head is as likely seen 
as the outcome tail. But what would be the average? There is no way of 
answering this question using common language -- for sure the coin does not 
have a `Cecrops' 
face that is half head and half tail. To make any statistical consideration such 
as taking the average, it is necessary to assign numerical values to outcomes 
of chance, called a random variable, so for instance we can associate the 
number 1 to the outcome head, and 0 to the outcome tail. We can then 
meaningfully say that the average of the outcome of a fair coin is 0.5 without 
any difficulty. In a similar vein, to model decision making under uncertainty 
it is necessary to introduce a random variable to represent different options, 
and likewise another random variable to represent noise. The idea of 
assigning numerical values to rumours, speculations, estimations, news 
bulletins, etc., may 
appear rather abstract, and it requires another leap in imagination to realise 
that this is in fact no more abstract than associating the values 0 and 1 to the 
outcomes of a coin tossing. Indeed, the variable $X$ in a 
decision making refers to the correct choice. The information $\xi=X+\epsilon$ 
thus does not refer to observing one's own decision process. Rather, $\xi$ 
embodies the observation of external information sources in relation to 
arriving at the correct choice in the decision making.  

The example above in which the observation is characterised by the relation 
$\xi=X+\epsilon$ is, of course, meant to represent the simplest situation, 
whereas in real-life decision makings, the noise typically changes in time and 
is thus represented by a time series $\{\epsilon_t\}$, where $t$ denotes the 
time variable. In some cases, the 
nature of available decision options itself may change in time, in which case 
$X$ will also be replaced by a time series. At any rate, in almost all realistic 
circumstances, the information-providing observation is not a fixed random 
variable, but rather is given by a time series $\{\xi_t\}$. Fortunately, the theory 
of signal detection and communication is highly developed 
(Davis 1977, Liptser \& Shiryaev 2001) so as to allow for 
a good level of tractability to model many realistic circumstances in decision 
making. 

\section{Modelling electoral competition with information}

An information-based approach to modelling the dynamics of electoral 
competitions has been introduced recently in Brody \& Meier (2022) and in 
Brody (2019). The idea can be sketched as follows. In the context of an 
electoral competition, a voter typically has a handful of issues of concern 
(such as taxation policy, climate policy, education policy, policies on abortion 
right and gun control, or perhaps the personality of the leader of a political 
party, \&c.), and likewise possesses partial information about how different 
candidates, if elected, will implement policies to address these issues. Each 
such issue is then modelled by a random variable so as to assign numerical 
values to policy positions, and these random variables, 
whose values represent different policy positions different candidates would 
implement, play the role of signals whose values the voters wish to identify.  
Hence for example in the case of a binary issue 
(for or against), one can assign, say, the values $+1$ and $-1$ to the two 
positions. Each such random variable is referred to as a `factor' and for each 
electoral 
factor there is a noisy observation characterised by a time series. Thus 
voters can only infer the best estimates for the values of these factors, based 
on available information. 

For a given voter, their preferences on different policy positions are then 
modelled by weights $\{w_k\}$, which are not necessarily positive numbers. 
The signs of the weights reflect their preferences on the 
various issues, while the magnitude $|w_k|$ represents 
the significance of the policy position about the $k$-th issue for that voter. In 
Brody \& Meier (2022) a linear scoring rule was assumed to associate for 
each candidate a score from a given voter, determined by the weighted 
average of their best estimates for different factors. That is, 
the score $S_l$ assigned to candidate $l$ by a voter with preference pattern 
$\{w_k\}$ is given by 
\[ 
S_l = \sum_{k} w_k \, {\mathbb E}[X_k] ,
\]
where ${\mathbb E}[-]$ denotes expectation operation. 
A given voter will then 
choose to vote for the candidate with the highest score. 
The importance of imperfect information about the policy positions of the 
candidates in electoral competitions has been noted before (see, eg.g, 
Harrington, 1982; McKelvey, \& Ordeshook 1985;  Feddersen \& 
Pesendorfer 1997; Fowler \& Margolis 2014). The approach of Brody \& 
Meier (2022) is to take this idea a step further by modelling the noisy flow 
of information concerning the values of the policy positions in the form of 
a time series $\{\xi_t\}$, from which 
the dynamics of the opinion poll statistics can be deduced. This is because 
the expectation ${\mathbb E}[-]$ 
is now replaced by a conditional expectation subject to the 
noisy information flow regarding the policy positions of the candidates. 

Another advantages of this approach, apart from being able 
to simulate the time development of the conditional expectations of the 
electoral factors $\{X_k\}$, is that given the information about 
the distribution of voter preferences within a group of the population, it is 
computationally straightforward to sample a large number of voter profiles 
(the weights $\{w_k\}$) without going through the costly and time-consuming 
sampling of the actual voters. Thus, for example, if there 
were one million voters, and if we have the knowledge of the distribution of 
voter preferences on different issues, then by sampling from this distribution 
a million times we can artificially create voter preference patterns, from which 
we are able to simulate the dynamics of the opinion poll statistics and study 
their implications.
As a consequence, the information-based 
approach makes large-scale simulation studies and scenario analysis on 
behavioural pattern feasible, when it comes to systems driven by information 
revelation under uncertainties.  

It should be evident that because the starting points of the formalism based 
on communication theory 
are (a) to identify relevant issues and associate to them random variables, 
called factors, 
and (b) to build a model for the flow of information for each of the factors, it 
readily suggests a way to explore how adjustments of information flow (for 
example, when to release certain information) will affect the statistics of the 
future (such as the probability of a given candidate winning on the election 
day). Furthermore, it also suggests a way to model deliberate disinformation 
and study their impacts. These ideas will be explained in 
more detail below. 

\section{Disinformation and their impacts}

The intention of deliberate disinformation -- the so-called `fake news' -- is, 
as many people interpret the phrase nowadays, to create a bias in people's 
minds so as to impact their behaviours and decision makings. But clearly 
such disinformation will have little impact if the person who 
receives the information is aware of this. That is, if the person has an 
advanced knowledge of the facts, then they will not be impacted by false 
information -- although there are suggestions that there can 
be such `anchoring' effect even among well-informed individuals 
(Wilson. \textit{et al}. 1996). (The situation is different if a 
false information is given first, and the truth is revealed subsequently, 
because in this case the prior belief has been shifted before the facts are 
revealed.) Because disinformation is not part of the 
`signal' that in some sense represents truth, it can be viewed as 
a kind of noise. In the context of a traditional communication channel, on the 
other hand, while noise is a nuisance, it does not have an intent. In other 
words, noise does not have an unknown bias. Putting these together, it should 
be apparent that the release of deliberate disinformation is equivalent to the 
introduction of a bias into the noise that is undetected by the receiver of the 
information. Thus, in the case of the earlier example, the observation under 
disinformation takes the 
form 
\[ 
\eta = X + (\epsilon + f) , 
\] 
where $f$ represents the biased disinformation so that the expectation of 
$f$ is nonzero. If a 
person receives the value of $\eta$, but is unaware of the existence of 
the term $f$ and presumes that it is the value of $\xi=X+\epsilon$, then the 
resulting inference will be misguided. For instance, a positive $f$ will 
misguide people in thinking that the value of $X$ is larger than what it 
actually is, and conversely for a negative $f$ people will be misled to 
the conclusion that the value of $X$ is smaller than what it actually is. 
Again, once one recognises the need 
for the introduction of a random variable $f$ for representing disinformation 
so as to allow for a meaningful statistical treatment, it becomes apparent how 
to model and study behavioural dynamics in the presence of fake news. 

Continuing on with this simple example, where $X$ is a binary random variable 
with \textit{a priori} probabilities $(p,1-p)=(0.5,0.5)$ and $\epsilon$ is a 
zero-mean normal random variable with standard deviation $\nu=0.2$, 
suppose that disinformation is released so as to enhance the probability that 
the choice corresponding to $X=0$ is selected by a decision maker. The 
decision maker is under the assumption that the observation is of the form 
$\xi=X+\epsilon$. This means, in particular, that the smaller the value of $\xi$ 
is, the higher the \textit{a posteriori} probability of $X=0$ is. To enhance the 
\textit{a posteriori} probability, suppose, in the previous scenario whereby 
$\xi=0.73$, that the released disinformation amounts to the realisation that 
$f=-0.093$. Then the perceived, or deceived \textit{a posteriori} probability is 
$p'\approx0.032$, even though in reality the number ought to be 
$p'\approx0.0032$. 

In the above example, the disinformation-induced perceived \textit{a 
posteriori} probability, although has been enhanced by a factor of ten, 
remains too small to be of significance in affecting decision making. 
However, it has to be recognised that in reality the information flow is 
typically continuous in time, i.e. for real-world applications 
to modelling behavioural dynamics of the public one has to be working with 
a time series rather than a single-shot information model considered here. 
What this means is that while each disinformation may only shift the public 
perception by a small amount, the impact of a relentless release of 
disinformation accumulates in time to become significant.  

\begin{figure}[t]
      \centering
        \includegraphics[width=0.47\textwidth]{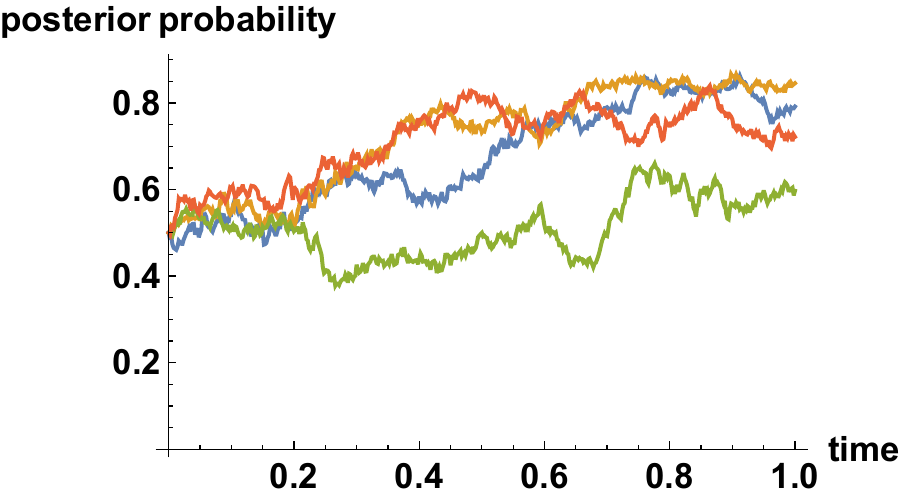} 
        \includegraphics[width=0.47\textwidth]{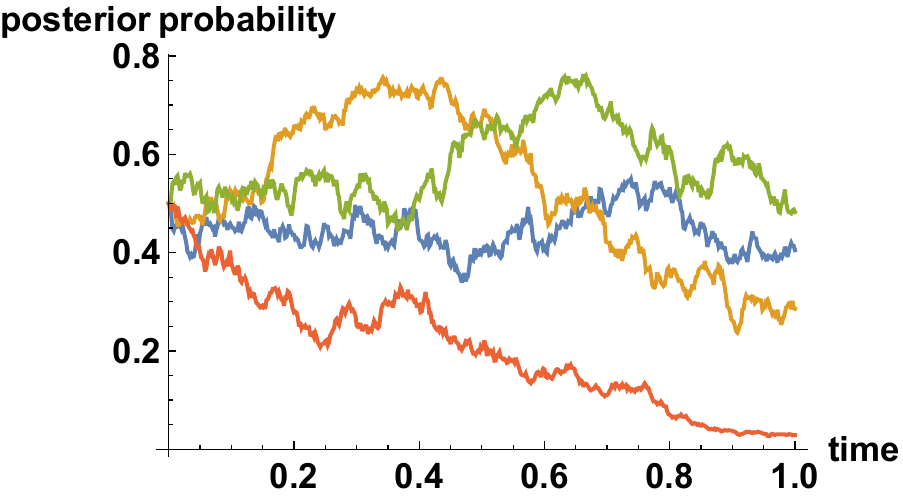}
        \caption{\textit{Impact of fake news}. 
        The two alternatives are represented by the values $X=0$ and 
        $X=1$, but at time zero the uncertainty is at its maximum (the opinion 
        is equally split between the two alternatives). Then noisy observation 
        begins, and the prior opinion is updated in time 
        (expressed in years). Plotted 
        here are sample paths for posterior probabilities that $X=1$. 
        In all simulations the 
        correct choice is (secretly) preselected to be the one corresponding 
        to $X=1$. Depending on how the Brownian noise develops the best 
        inference develops differently, as shown here on the left panel, but by 
        waiting sufficiently long, ultimately enough truth is learned and all 
        decisions on the left panel will converge to the correct one, 
        typically by time $t\approx10$.  
        The impact of disinformation, 
        released at a random point in time (at $t=0.6$ here), is shown on the 
        right panel. The disinformation is released at a constant rate in 
        time, intended to mislead the decision maker to select the choice 
        corresponding to $X=0$. The rate of release is taken to be sufficiently 
        strong that if decision makers are unaware of the disinformation 
        then they will be led to making the incorrect decision ($X=0$), even 
        though the simulator had preselected the correct decision to be that 
        corresponding to $X=1$.} 
        \label{fig_1}
 \end{figure}

To visualise the effect, consider a time-series version of the model in which 
the time-series $\{\epsilon_t\}$ for noise is represented by a Brownian motion 
(hence for each increment of time the noise is normally distributed with mean 
zero and variance equal to that time increment), but the signal $X$ remains a 
zero-one binary random variable, whose value is revealed at a unit rate in 
time. Thus the observed time series takes the form 
\[
\xi_t = X t + \epsilon_t 
\]
in the absence of disinformation, whereas the Brownian noise 
$\{\epsilon_t\}$ acquires a drift term $f(t-\tau)$ at some point $\tau$ 
in time in the 
presence of disinformation. 
Example sample paths with and without a release of fake news are 
compared in Fig.~\ref{fig_1}. 

One of the advantages of the present approach is that a simulator can 
preselect what is ultimately the `correct' decision. Looking at each realisation 
one cannot tell, without waiting for a sufficiently long time, which way the 
correct decision is going to be. Nevertheless, the simulator is able to select 
the correct decision in advance and let the simulation run. In this way, a 
meaningful scenario analysis can be pursued. With this in mind, in 
Fig.~\ref{fig_1} sample paths are shown, all of which corresponds to the 
realisation that the decision corresponding to the value $X=1$ is ultimately 
the correct decision. On the left panel, starting with a 50-50 prior opinion, 
the development of the posterior opinion based on the observation of the 
time series $\{\xi_t\}$ is shown for four different realisations of the noise 
$\{\epsilon_t\}$. Depending on how the noise develops, the realisations 
will be different, but in all cases, ultimately, by waiting longer than the 
timescale shown here, the correct decision (selected by the simulator) will 
be selected by the decision makers. In contrast, if sufficiently strong 
disinformation intended to guide decision makers towards the incorrect 
choice (we know that it is incorrect because the simulator did not choose 
that decision) is released at some point in time, and if nothing is done 
about this so that decision makers are unaware of this, then 
ultimately all decisions will converge to the incorrect choice, as shown on 
the right panel.  

It is worth remarking in this connection that in real-world 
applications there are two situations that arise: One in which the correct 
decision will be revealed at some point, and one in which this is never 
revealed. For instance, if the decision is whether to invest in asset $A$ or 
asset $B$ over an investment period, then at the end of the period one 
would discover which investment resulted in a higher yield. For other 
decisions, for example in the context of an election or referendum, it is often 
the case that voters will never find out, beyond an informed guess in some 
cases, which candidate would have been better, because the lost candidates 
do not have the opportunities to demonstrate the outcomes of their policy 
implementations. In fact, even for the winning candidate, the merits of their 
policies may not become apparent, especially when there are long-term 
consequences. These latter cases amount to having a signal that drops 
out at some point in time (e.g., on the election day), and 
hence afterwards the public no longer 
receives information to improve their assessments. Communication theory 
allows for the flexibility to handle all these different situations that 
might arise in reality. For example, if the correct choice is 
revealed for sure after a finite time horizon, then one can let the noise $\{
\epsilon_t\}$ be modelled by a Brownian bridge process that goes to zero at 
the end of the period (Brody \textit{et al}. 2022). Alternatively, if the correct 
choice is never revealed, then one can let the information revelation rate 
$\sigma$ introduced below to vanish at some point in time. 

\section{Information control}

Besides the impact of disinformation, there is another 
important ingredient that has to be brought into the analysis when considering 
the controlling of public behaviour. This concerns, for instance, a situation in 
which there are individuals who are aware of the value of $X$ that the public 
at large are trying to infer. In such a situation, what one might call the 
information-flow rate, or the signal-to-noise ratio, may be adjusted. To 
understand this, let us return to our single-shot information model, but this 
time we have 
\[ 
\xi = \sigma X + \epsilon , 
\] 
where the parameter $\sigma$ determines the magnitude of the signal. To 
understand the effect of $\sigma$, let us take an extreme case where 
$\sigma=100$ while $X$ is a zero-one binary variable and $\epsilon$ is a 
zero-mean normal variable with a small standard deviation. Then for a given 
value of the noise $\epsilon$ there are two possible observations: $\xi=
\epsilon$, or $\xi=100+\epsilon$. Because the realised value of $\epsilon$ 
will almost certainly be close to zero, we know already that $\xi\approx0$ if 
$X=0$ and $\xi\approx100$ if $X=1$. Hence the effect of $\sigma$ is to 
amplify the signal, making it easier to infer the value of $X$. 
Conversely, suppose that $\sigma=0.01$ in this example. Then we know that 
$\xi\approx0$ irrespective of whether $X=0$ or $X=1$. Hence the observation 
will be of little help in inferring the value of $X$: the signal is dimmed by having 
a small value of $\sigma$. 

With this example in mind it should be evident that the general information 
model can take the form 
\[ 
\eta = \sigma X + (\epsilon+f) .
\] 
To control the behaviour of the public, one can either introduce the term 
$f$ with a nonzero mean in such a way that the public is unaware of its 
existence, and hence confuses the contribution of $f$ as arising from $X$, or 
increase (decrease) the value of $\sigma$ so that the public can arrive at 
a more reliable inference faster (slower). These are the two fundamental 
ways in which the public behaviour under uncertain environments can be 
manipulated via information. 

It is worth remarking here, incidentally, that if $f$ has no bias, then its effect 
is equivalent to reducing the value of $\sigma$, because in this case one is 
merely enhancing the magnitude of noise. Hence the introduction of purely 
random disinformation has the effect of slowing down the public from 
discovering the truth. This may be intuitively apparent, but here it follows as a 
direct consequence of communication theory. In particular, in the context of 
an observation involving a more general time series, the timescale of 
arriving at a reasonable inference about the value of $X$ is typically 
proportional to $\sigma^{-2}$. This is the timescale for which the amount of 
uncertainty as measured by the variance of $X$ is reduced by 
more than 50\% of the initial uncertainty. Hence if the magnitude of noise is 
doubled, then it takes four times longer to arrive at the same level of 
inference. 

With the above characterisation of the two fundamental ways in which 
information can be manipulated, it is possible to ask which strategy maximises 
the chance of achieving a certain objective, and techniques of communication 
theory can be used to arrive at both qualitative and quantitative answers. As 
an example, consider an electoral competition, or a referendum. To simplify the 
discussion let us assume that the choice at hand is binary, and the information 
providing process is a time series, where both the noise 
$\{\epsilon_t\}$ and the information revelation rate $\{\sigma_t\}$ are changing 
in time. If an agent is willing to engage in a 
strategy to divert the public to a particular outcome based on disinformation, 
then the example illustrated in Fig.~\ref{fig_1} shows that it suffices to release 
`fake news' whose magnitude $|f_t|$ is greater than that of the information 
revelation rate $|\sigma_t|$. However, there are two issues for the fake-news 
advocators: First, the strategy is effective 
only if the public is unaware of the existence of disinformation. Some people 
are knowledgable, while others may 
look it up or consult fact-checking sites. From these, some 
can infer the probability distribution of disinformation, even though they may 
not be able to determine the truth of any specific information, and the 
knowledge of this distribution can provide a sufficient deterrence against the 
impact of disinformation (Brody \& Meier 2022). Second, a frequent release 
of information can be costly. For a state-sponsored disinformation unit this may 
not be an issue, but for most others, it is typically rather costly to disseminate 
any information to the wider public -- for example, by paying a lot of money to 
the so-called `influencers' to discourage people from being vaccinated against 
a potentially deadly virus. 

From the viewpoint of a fake-news advocator, the cost issue can be addressed 
by means of signal-processing techniques outlined here. For instance, 
suppose that for cost reasons there is only one chance of releasing 
disinformation, whose strength grows initially but over time is damped down, 
perhaps owing to people discovering the authenticity of the information. 
In such a scenario one would be interested in finding out 
the best possible timing to release disinformation so as to maximise, for 
instance, the 
probability of a given candidate winning a future election. The answer to such 
a question of optimisation can be obtained within the present approach 
(Brody 2019).

\begin{figure}[h]
      \centering
        \includegraphics[width=0.47\textwidth]{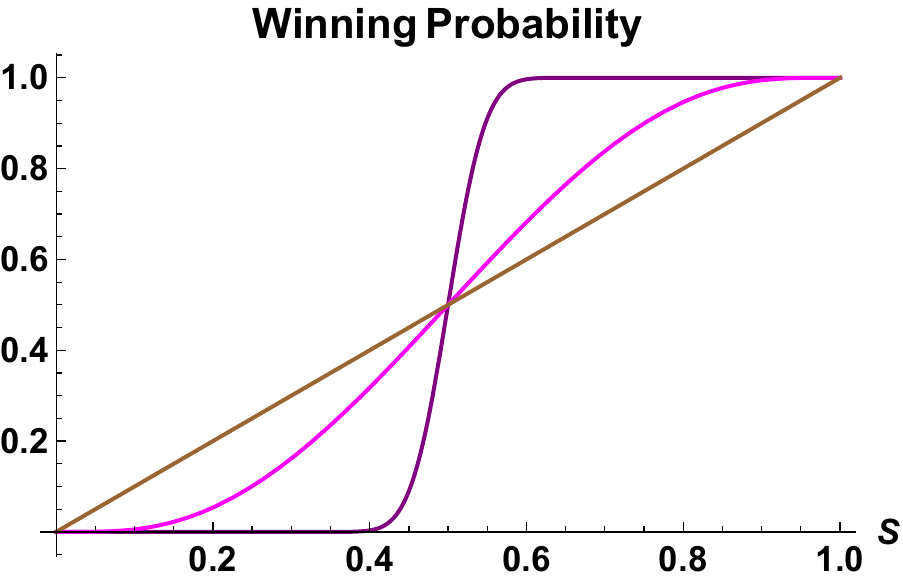} 
        \includegraphics[width=0.47\textwidth]{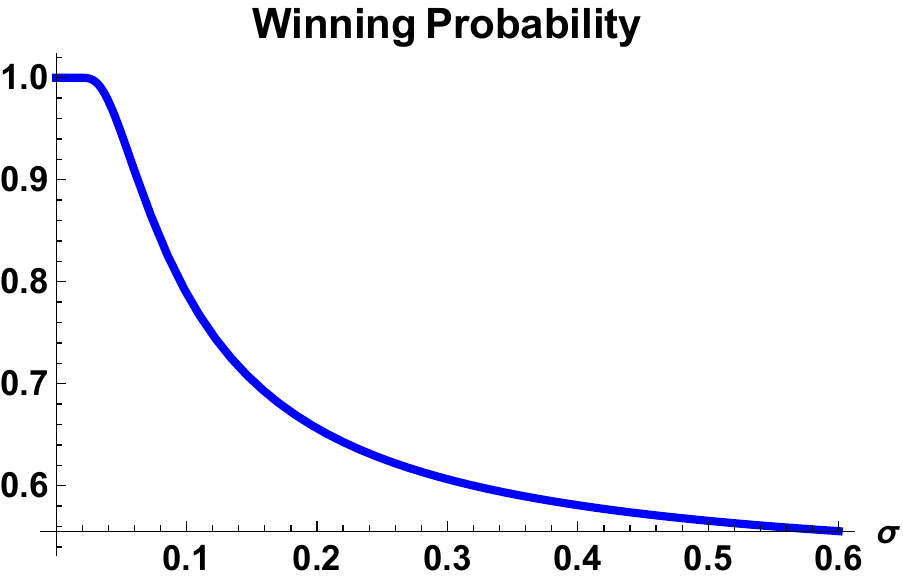}
        \caption{\textit{Probability of winning a future election}. 
        The winning probability of a candidate in a two-candidate electoral 
        competition, to 
        take place in one year time, is plotted. On the left panel, the 
        probabilities are shown as a function of today's support rating $S$ for 
        two different values of the information-flow rate: $\sigma=0.15$ 
        (purple) and $\sigma=0.95$ (magenta). If today's poll $S$ were an 
        indicator for the winning probability, then it would be given by a straight 
        line (brown), but in reality the probability of winning a future election 
        of a candidate, whose current support rate is $S>50\%$, is always 
        greater than $S$. On the right panel, the winning probability is shown 
        as a function of the information-flow rate $\sigma$ of a candidate 
        whose support rate today is $S=52\%$. If the candidate is leading the 
        poll, then the best strategy is to reveal as little information relevant to 
        the election as possible. } 
        \label{fig_x2}
 \end{figure}

From the viewpoint of an individual, or perhaps a government, who wishes to 
counter the impact of disinformation, on the other hand, the analysis presented 
here will allow for the identification of optimal strategies potentially adopted by 
fake-news advocators so as to anticipate future scenarios and to be prepared. 
It also provides a way for developing case studies and impact analysis. This is 
of importance for two reasons. First, the conventional approach to counter the 
impacts of fake news, namely, the fact checking, although is an indispensable 
tool, does not offer any insight into the degree of impact caused by fake news.  
Second, while information-based approach tends to 
yield results that are consistent with our intuitions, some conclusions that can 
be inferred from the approach are evident 
with hindsight but otherwise appear at first to be counterintuitive. Take, 
for instance, the probability of a given candidate winning a future election, 
in a two-candidate race, say, candidates $A$ and $B$. It can be shown 
(Brody 2019) that 
if the current poll suggests that candidate $A$ has $S\%$ support, then the 
actual probability of candidate $A$ winning the future election 
is always greater than $S$ if 
$S>50$, and is always less than $S$ if $S<50$. Hence, contrary to a naive 
intuition, the current poll statistics are not the correct indicators for the actual 
realised probabilities of election outcomes. Further, the smaller the 
information flow rate $\sigma$ is, the greater is the gap between the current 
poll and the realised winning probability. Thus, for instance, if $S=51$, say, six 
months before the election day, and if the value of $\sigma$ is very small, then 
the projected probability of candidate $A$ winning the election is significantly 
higher than $51\%$, and in the limit $\sigma$ tending to zero, it approaches 
$100\%$ (see Figure~\ref{fig_x2}).

This may at first seem counterintuitive, but with reflection one can understand 
why this has to be the case. If the value of $\sigma$ is close to zero, then what 
this means is that virtually no information about the factor $X$ will be revealed 
between now and the election day six months later (here it is assumed for 
simplicity of discussion 
that there is only one factor). But without reliable information people 
do not change their mind spontaneously. Hence if candidate $A$ has $51\%$ 
support today, then without further information 
$51\%$ of voters will continue to support candidate $A$ 
six months later, meaning that the actual probability of winning is closer to 
$100\%$. It follows that if a candidate is leading the poll, then it is in their best 
interest not to reveal any information about their policies or personality, 
unless there are good reasons in doing so to further enhance the current lead. 
Conversely, if a candidate is lagging behind the poll then it is in their best 
interest to reveal as much information as possible, so as to create movements 
that may change the balance of the poll. 

This example naturally lends itself to the second way in which information 
can be controlled. Namely, to adjust the value of $\sigma$. This is a different 
approach from the one based on releasing disinformation to guide people 
away from discovering facts. For example, if there is a fact, such as tax return, 
that a candidate does not wish the public to find out, or if a candidate is 
leading the poll statistics even though the candidate has no clue about future 
policies, then the value of $\sigma$ can be reduced either 
by not revealing any information or simply by putting out a lot of random noise 
peripheral to the issue. Alternatively, if the value of $X$ is known to a small 
number of individuals (e.g., the candidates themselves) when it is 
advantageous for the candidate that the public should discover this, then 
they are in the 
position to release more information to enhance the value of $\sigma$. In a 
more general situation where $\{\sigma_t\}$ is time dependent, it is possible 
to design how the information revelation rate should be adjusted in time 
(Brody 2019) so 
as to maximise the objective (for example, maximising the probability of 
winning an election, or maximising the sales in the context of advertisements).

\section{Information clusters and tenacious Bayesians}

One of the key issues associated with the deliberate dissemination of 
disinformation in a coordinated and organised manner (for example, by a 
state-sponsored unit) concerns the fact that although there is a very wide 
range of information sources readily available, people have the tendency of 
gathering information from a limited set of sources, resulting in the 
creation of clusters of people digesting similar information, and this can be 
exploited by a malicious fake-news advocator. To understand the formation 
of such clusters, consider the following scenario in a different context. 
Imagine that there is a wide open space, with a large number of people 
standing at random, and that these people are instructed to lie down in such 
a way that they lie as parallel as possible with their neighbours. Or alternatively, 
the instruction may be that everyone should lie with their heads pointing either 
north or south, such that they should lie in the same orientation as their 
neighbours. In theory, there 
are alignments such that all the people lie in a perfectly parallel configuration 
(for instance, they all lie with their heads pointing north), but such a 
configuration will not be realised in reality. The reason is because the 
instruction that they should lie as parallel as possible with their neighbours 
is a local one, and a local optimisation does not yield a global optimisation 
when there is a wide-ranging complex landscape of possible configurations. 
As a consequence, what will happen is the formation of vortices or clusters, 
in the latter case separated 
by domain walls separating alignment mismatch, 
where within a cluster people are closely aligned. 

Formation of informational clusters are perhaps not dissimilar to this. The 
highly developed nature of Internet might give the impression that everything 
is `global' in this information society, but this is not the case because the 
concept of a neighbour in an information cluster, where 
people within a cluster digest similar information sources, 
 need not have any relation 
to a geographical neighbour: a person living across the Atlantic can be a 
neighbour in the information cluster, while the next door occupant can be 
from another universe for that matter. As a consequence of the cluster 
formation, the type of information digested in one cluster tend to differ from 
that in another cluster. For instance, a regular reader of a left-leaning news 
paper is unlikely to pick up a right-leaning paper, and \textit{vice versa} -- 
the heights of the domain walls are made higher by the fluidity of Internet, 
and, in particular, by fake news. 

Of course, those belonging to a given cluster are often well aware of the 
existence of other 
opinions shared by those in other clusters. Yet, those counter opinions -- 
the so-called `alternative facts' -- have seemingly little impact in altering 
people's opinions, at least in the short term. The reason behind this can be 
explained from a property of Bayesian logic. Indeed, one of the consequences 
of the clustering effect is the tendency of placing heavier 
prior probabilities on positions that are shared by those within the cluster. 
The phenomenon of overweighting the prior is sometimes referred to as 
`conservatism' in the literature (El-Gamal \& Grether 1995), although this 
terminology can be confusing in a political discussion. At any rate, when 
the prior probability is highly concentrated towards one of the alternatives, 
and if this happens to be ultimately the `incorrect' choice, then even if 
counter facts are presented time and again, the prior weight need not 
change very much for a long time under the Bayesian inference. This 
phenomenon will be referred to as the `tenacious Bayesian inference' here. 

\begin{figure}[t]
      \centering
        \includegraphics[width=0.47\textwidth]{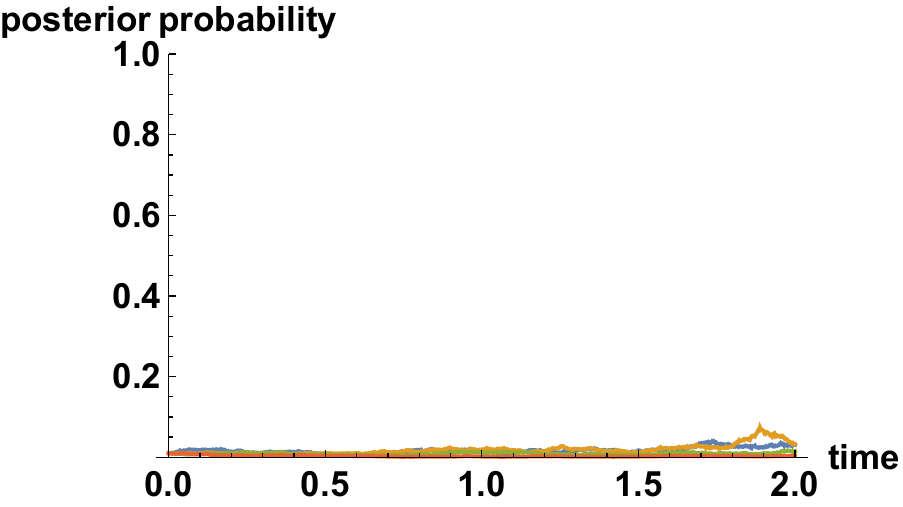} 
        \includegraphics[width=0.47\textwidth]{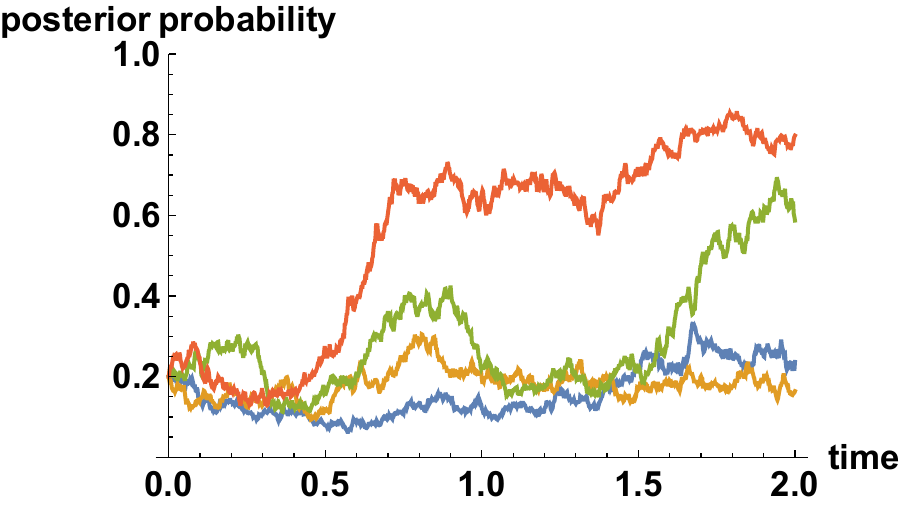}
        \caption{\textit{Tenacious Bayesian behaviour in a binary decision making}. 
        The two alternatives are represented by the values $X=0$ and 
        $X=1$, and in all simulations, the `correct' decision is preselected 
        to correspond to the choice $X=1$. On the left panel, four sample 
        paths for the a posteriori probability that $X=1$ 
        are shown when the \textit{a priori} probability for the incorrect 
        decision $X=0$ is given by 99\%. Although the evidences presented 
        by the observed time series consistently indicate that $X=0$ is not 
        the correct choice, the decision makers in these simulations have 
        hardly 
        changed their views even after two years, 
        in spite of following the Bayesian logic. 
        In contrast, if the prior probability for the incorrect choice $X=0$ is 
        reduced to, say, 80\%, then with the same amount of 
        information-revelation rate ($\sigma=1$ in both cases) 
        there will be more variabilities, as shown on the right panel.} 
        \label{fig_2}
 \end{figure}
 
The mechanism behind the tenacious Bayesian phenomenon can be explained 
by means of communication theory. It has been remarked that for the 
uncertainty to reduce on average to a fraction of the initial uncertainty, a 
typical timescale required for gathering information is proportional to the 
inverse square of the information flow rate $\sigma$. More precisely, the 
timescale is given by $(\sigma\Delta)^{-2}$, where $\Delta^2$ is the initial 
uncertainty, measured by the variance. Hence if the prior probability is 
highly concentrated at one of the alternatives, then $\Delta$ is very small, 
so typically it will take a very long time for the initial uncertainty to reduce by 
a significant amount. This is not an issue if the initial inference is the correct 
one. However, if the initial inference is incorrect, then there is a problem, for, 
the uncertainty will have to significantly increase before it can decrease again. 
As a consequence, having a very high prior weight on any one of the 
alternatives means it is difficult to escape from that choice even if ultimately 
it is not the correct one, because each alternative acts like an attractor. 
Sample paths illustrating this effect are shown in Fig.~\ref{fig_2}.  

In the characterisation of human behaviour it is sometimes argued that people 
act in an irrational manner if they do not follow the Bayesian rule. So for 
instance if a person is presented with a fact that diametrically contradicts their 
initial view, and if the person does not change their view afterwards, then this 
is deemed counter to Bayesian and hence irrational. While it is not 
unreasonable to associate irrationality with a lack of Bayesian thinking, any 
experimental `verification' of irrational behaviour based on this criterion is 
questionable, due to the tenacious Bayesian phenomenon. 
A good example can be seen in the 
aftermath of the 2020 US presidential election. Many believed (and still do) 
that the election outcomes were `rigged' even though the large number of 
lawsuits put forward challenging the outcomes were thrown out of courts one 
after another. Although the factual evidences presented suggested that the 
election results were not rigged, this had little influence on those who 
believed the contrary. One might be tempted to argue that this behaviour is 
irrational, but a better characterisation seems to be that these people are 
acting rationally in accordance wth their Bayesian logic, albeit they have 
strongly skewed priors. 
 
It should be evident that the effect of fake news naturally is to exacerbate 
the issue associated with the concentration of prior weights on incorrect 
inferences. In particular, if the prior weight for an incorrect inference is 
already high, then it does not require a huge amount of disinformation to 
maintain this status. Therefore, the phenomenon of tenacious Bayesian 
behaviour will have to be taken into account in exploring measures to 
counter the impacts of fake news. 

One immediate consequence of the tenacious Bayesian 
behaviour is that it explains, at least in part, the confirmation bias within 
the Bayesian logic. Broadly speaking, confirmation (or confirmatory) bias 
refers to a well-documented behaviour whereby people with particular 
views on a given subject tend to interpret noisy information as confirming 
their own views (Klayman 1995, Nickerson 1998, Martindale 2005). 
Thus, two people with opposing views, when 
presented with the same ambiguous information, may simultaneously 
interpret the information as supporting their initial views. If, in particular, 
the polarisation of the two opposing views increases after digesting the 
same noisy information (Lord, \textit{et al}. 1979), then this is considered 
as a clear evidence that people do not follow Bayesian thinking 
(Griffin \& Tversky 1992, Rabin \& Schrag 1999, Dave 
\& Wolfe 2003, Nishi \& Masuda 2013, Rollwage \& Fleming 2021). 

The tenacious Bayesian behaviour observed here, however, 
suggests that such a phenomenon is not necessarily incompatible with the 
Bayesian logic, and hence that, contrary to common assertion, to a degree, 
confirmation bias can be explained as a consequence of Bayesian thinking. 
To establish that the tenacious Bayesian behaviour is a generic feature of 
Bayesian updating under uncertainties, it is necessary to work directly within 
the state space of decision making, which will be explained now. 

Suppose that the views held by decision maker $A$ on 
a set of $n$ alternatives is represented by the probabilities $(p_1, p_2, 
\ldots,p_n)$, while that of decision maker $B$ is represented by $(q_1, 
q_2, \ldots,q_n)$. To determine the level of affinity it will be useful to 
consider instead the square-root probabilities $\psi_i=\sqrt{p_i}$ and 
$\phi_i=\sqrt{q_i}$. These square-root probabilities then represent the 
\textit{states of decision makers}. More specifically, the state space of 
decision making is a vector space of unit-normalised positive vectors 
endowed with the Euclidean inner product, such 
that squared components of the vector determine probabilities for 
different alternatives. The separation of the two decision 
makers $A$ and $B$ can then be measured in terms of the spherical 
distance 
\[ 
\theta = \cos^{-1} \left( \sum_{i=1}^n \psi_i \, \phi_i \right) ,
\]
known in statistics as the Bhattacharyya distance (Brody \& Hook 2009). A 
definite state 
of a decision maker is represented by elements of the form $e_k = (0,\ldots, 
0, 1, 0, \ldots, 0)$, where only the $k$-th element in $e_k$ 
is nonzero. If two decision 
makers have identical views, then their separation distance vanishes, while 
if the distance takes its maximum value $\theta=\pi/2$ then their views are 
orthogonal, and hence incompatible. 

If the vector $\{\psi_i\}$ represents the prior state of decision maker $A$, 
and if noisy information relevant to the choice is revealed, then the prior 
will be updated to a posterior state $\{\phi_i\}$ in accordance with the 
Bayes formula, in the sense that the transformation 
$\psi_i^2\to\phi_i^2$ is determined by the Bayes formula. 
Now in the continuous-time setup where the noise is modelled by a 
Brownian motion, it is known in signal detection that the 
transformation of the posterior probability is governed by
a differential equation known as the Kushner equation (Kushner 1964). 
Translating this equation into the state-space by use of the square-root map, 
one finds that the deterministic component (the drift) of the dynamics is given 
by the negative gradient of the variance of the signal that is to be inferred from 
the noisy information (cf. Brody \& Hughston 2002). 
The nature of a negative gradient flow 
is to push the state into another state of a lower variance. 
What this means is that if the state of a decision maker 
is close to one of the definite states, say, $e_k$, for which the variance is 
zero, then the flow generated by Bayesian updating has the tendency of 
driving the state towards $e_k$. Putting the matter differently, the definite 
states $\{e_i\}$ are the attractors of the Bayesian flow. 

Now the variance is a measure of uncertainty, so this feature 
of Bayesian flow is only natural: reduction of uncertainty is what learning is 
about, and this is the reason why Bayesian logic is implemented in many of 
the machine learning algorithms, since the Bayesian updating leads to 
maximum reduction in uncertainty. However, this attractive feature can also 
generate an obstacle in the context of decision making, because the prior 
view held by a decision maker is subjective and hence may deviate far away 
from objective reality. In particular, if the state of a decision maker is close to 
one of the false realities $e_k$, then the Bayesian flow will make it harder to 
escape from the false perception, although by waiting long enough, 
eventually a decision maker will succeed in escaping from a false attractor. Or, 
alternatively, if by a sheer luck the noise takes unusually large  values that 
take the state away from the attractor, then by chance a quick escape 
becomes possible, but only with a small probability. 

\begin{figure}[t]
      \centering
        \includegraphics[width=0.47\textwidth]{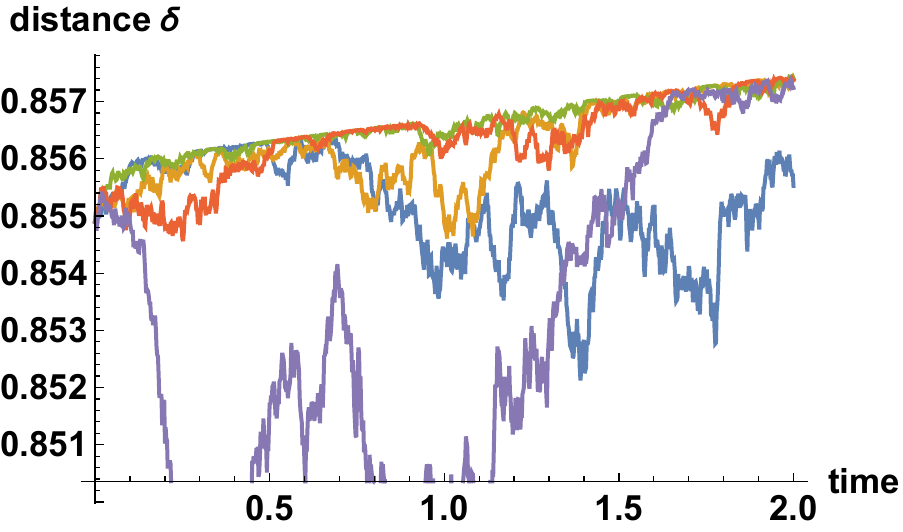} 
        \includegraphics[width=0.47\textwidth]{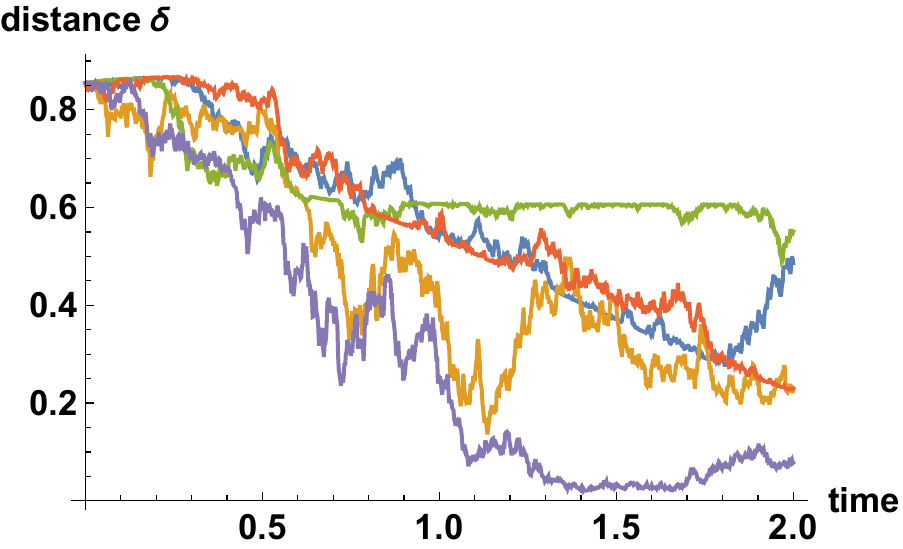}
        \caption{\textit{Separation distance under Bayesian 
        updating}. The polarity, or distance $\delta$ of two decision makers, 
        when they are 
        provided with an identical set of noisy information, has a tendency to 
        increase under the Bayesian updating, even though on average it 
        decreases in time. Five sample paths are shown here for two different 
        choices of $\sigma$. 
        On the left panel the information flow rate (signal to 
        noise ratio) is taken to 
        be $\sigma=0.2$. Simulation studies (not shown here) 
        indicate that in this case the upward 
        trend persists for some 40 years in about 50\% of the sample paths, 
        and the separation distance is typically reduced 
        to half of the initial value after about 100 years. When the information 
        flow rate is increased eleven-fold to $\sigma=2.2$, polarised 
        Bayesian learners 
        are forced to converge a lot quicker, as shown on the right panel, where 
        the separation is reduced to half of its initial value typically within two 
        years. } 
        \label{fig_z4}
 \end{figure}

With these preliminaries, let us conduct a numerical 
experiment to examine how the separation of two decision makers evolve 
in time under the Bayesian logic. Specifically, let there be five possible 
choices represented by a random variable $X$ taking the values $(1,2,3,
4,5)$. Decision maker $A$ assigns 96\% weight on the second alternative, 
whereas for other alternatives assigns 1\% weight each. Similarly, decision 
maker $B$ assigns 96\% weight on the third alternative, whereas for other 
alternatives assigns 1\% weight each. The initial separation of the two is 
thus given by $\theta\approx1.343$. Normalising the separation by setting 
$\delta=2\theta/\pi$ so that $0\leq\delta\leq1$ we find that the initial 
separation distance is given by $\delta_0\approx0.855$, 
where the subscript $0$ denotes the initial 
condition. Both decision makers are provided with the same noisy 
information represented by the time series $\xi_t=\sigma X t+\epsilon_t$, 
where the noise $\epsilon_t$ is modelled by a Brownian motion. The 
simulator can secretly preselect the `correct' decision to be, say, the fourth 
alternative so that both decision makers are trapped at wrong inferences. 
(The choice of the correct alternative will have little impact on the dynamics 
of the separation distance.) The results of numerical experiments are shown 
in Figure~\ref{fig_z4}. It should be stressed first that \textit{on average} the 
separation measure $\{\delta_t\}$ is a decreasing process, because 
Bayesian updating forces decision makers to learn. Yet, simulation study 
shows that there is a clear trend towards slowly increasing the separation 
measure over shorter time scales. That is, the separation tends to increase 
slightly, but when they decrease, the amount of decrease is more 
pronounced that on average it decreases. 

An important conclusion to draw here is that the separation 
of two decision makers \textit{can} increase under Bayesian inferences. 
Thus, while there is no suggestion here that confirmation bias can be fully 
explained by means of Bayesian logic, the analysis presented here, based 
on the tenacious Bayesian phenomenon, shows 
that the gap between empirical behaviours of people and those predicted by 
Bayesian logic in the context of confirmation bias is significantly smaller than 
what is often assumed in the literature.

\section{Psychology of false belief}

It is of interest to remark that methods of communication theory 
goes sufficiently far to allow for the simulation of an `alternative fact', 
that is, the simulation of an event whose probability of occurrence, or 
the \textit{a priori} probability perceived by the decision maker, is zero. 
In this connection it is worthwhile revisiting the meaning of the 
\textit{a priori} probabilities. In the context of natural science, these 
probabilities are interpreted to represent the objective probability of an 
event taking place. Thus, if an event with a very low \textit{a priori} 
probability were to occur, then the interpretation is the obvious one, 
namely, that a very rare event has occurred. In the context of social 
science, however, these probabilities need not characterise in any 
sense an objective reality. Hence, if a decision maker were to assign, say, 
a very low \textit{a priori} probability on one of the alternatives, then the 
interpretation here is that the probability merely reflects the subjective 
perception of that decision maker, while in reality the objective probability 
of that alternative being selected may remain high. In other words, a 
false belief does not represent a rare event. 

In an extreme case, a decision maker may assign zero probability to an 
alternative which may nevertheless represent reality. This can be 
viewed as an extreme limit of the tenacious Bayesian behaviour, except 
that, perhaps surprisingly, Bayesian logic here predicts that the 
psychology of a decision maker with a perfect false belief (that is, someone 
who assigns zero weight on an alternative that represents physical reality) 
exhibits an erratic indecisive behaviour different from the tenacious 
Bayesian characteristics. Such a behaviour is seen, however, only when 
there are more than two alternatives, for, if there are only two alternatives 
and if the \textit{a priori} probability is zero for one of them, then the view 
of the decision maker will not change in time under the Bayesian logic. 

\begin{figure}[t]
      \centering
        \includegraphics[width=0.47\textwidth]{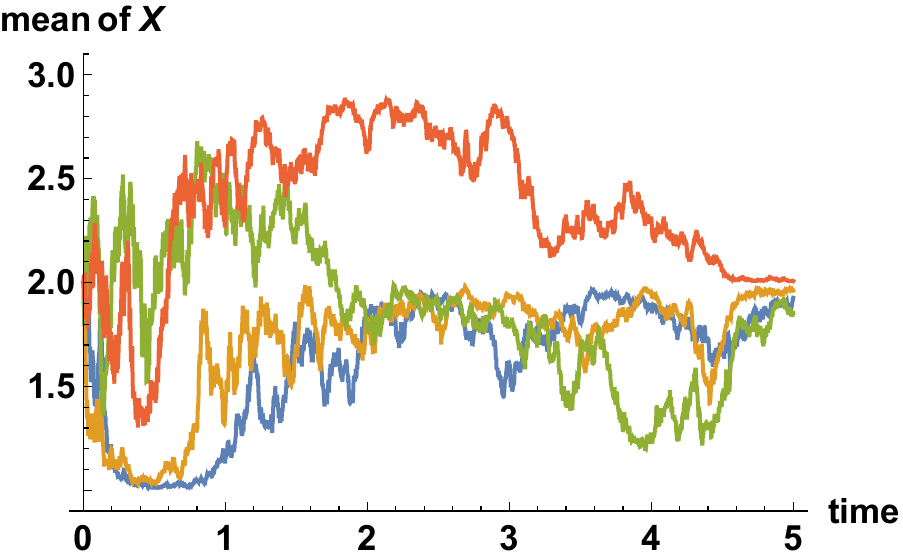} 
        \includegraphics[width=0.47\textwidth]{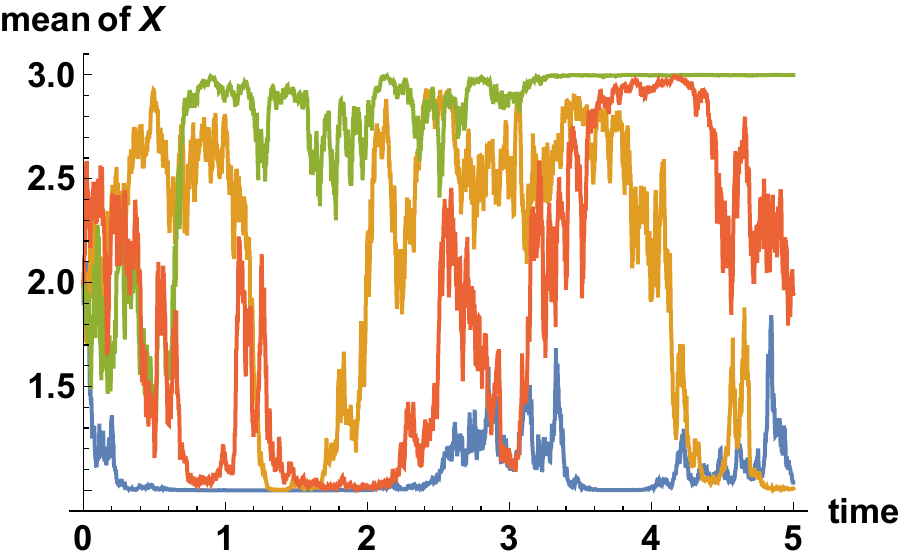}
        \caption{\textit{Simulating alternative fact}. 
        Three alternatives are represented by the values $X=1$, $X=2$, and 
        $X=3$. Plotted here are sample paths for the mean 
        values of $X$ subject to information process $\xi_t=\sigma X t + 
        \epsilon_t$, where $\sigma=2$ and $\{\epsilon_t\}$ denotes Brownian 
        noise. 
        In all simulations, the simulator has chosen the alternative 
        $X=2$ to be the correct one. On the left panel, all decision makers 
        start with the prior belief that the probability of $X=2$ is only 10\%, 
        whereas the two other alternatives are equally likely realised at 45\% 
        each. Hence the initial mean of $X$ equals $2$. 
        Initially, their views tend to converge either to $X=1$ or $X=3$; 
        but over time, sufficient facts are revealed that they all converge to 
        the correct choice made by the simulator. But what if they do not 
        believe in the `truth' at all? On the right panel, the simulator has again 
        chosen $X=2$ to represent the true value, but the decision makers 
        assume that this is impossible and that the two other alternatives are 
        equally likely realised at 50\% each. Hence again the 
        initial mean of $X$ equals $2$. In this case, the decision makers' 
        views tend to converge quickly to one of the two `false' alternatives 
        $X=1$ or $X=3$.  These two beliefs are, however, only quasi-stable; 
        the views will never converge indefinitely. Instead, their 
        views tend to flip back and forth between the alternatives $X=1$ and 
        $X=3$, but never converging to either one, and certainly never come 
        close to the correct alternative $X=2$. } 
        \label{fig_4}
 \end{figure}

Sample paths of such simulations are shown in Fig.~\ref{fig_4}, in the case where there are three alternatives. 
In general, when there are more than two alternatives, and if decision 
makers assign zero prior probability to the true alternative, then their views 
tend to converge quickly to one of the false alternatives, remain there for a 
while, before it jumps to one of the other false alternatives and then after a 
while jumps again to yet another false alternative. Such a `hopping' 
phenomenon can only be observed when they categorically refuse to 
believe in or accept 
the truth, and this erratic behaviour is predicted by Bayesian logic. 

The intuitive reason behind this hopping behaviour is that no false belief 
can ever be stable for long under the presence of information that reveals 
the reality. Hence the stronger is the information revelation rate about the 
reality, the more erratic the behaviour becomes. This feature can be 
studied alternatively by examining the Shannon-Wiener entropy (Wiener 
1948), which represents the measure of uncertainty. Under a learning 
process characterised by the observation of the noisy time series 
$\{\xi_t\}$, the uncertainty about different alternatives as characterised by 
entropy decreases on average. Hence a learning process is represented 
by the reduction of entropy, resulting in a low entropy state. 
This is why a decision maker who refuses to accept the real alternative 
quickly reaches a state of low entropy, and wishes to stay 
there. The reality however contradicts the chosen alternative. Yet, if 
entropy (hence uncertainty) were to now increase, even though the 
learning process continues, then this amounts to an admission of having 
to have rejected the truth. In other words, a state of high entropy is 
unstable in such a circumstance. 
The only way out of this dichotomy is to rapidly 
swap the chosen false alternative with another false alternative, until 
reality is forced upon the decision maker, at which point the second false 
alternative is discarded and replaced by either the original or yet another 
false alternative. This process will continue indefinitely. Only by a 
reinitialisation of the original assessment (for instance by 
a dramatic event that radically changes one's perception) 
in such a way that assigns a 
nonzero probability on the rejected alternative -- no matter how small 
that may be -- a decision maker can escape from this loop. 

It might be worth pondering whether the assignment of 
strictly vanishing probability (as opposed to vanishingly small probability) 
to an alternative by a decision maker represents a realistic scenario. 
Indeed, it can be difficult to determine empirically whether a decision 
maker assigns strictly zero probability on an alternative, although in 
some cases people seem to express strong convictions in 
accepting or rejecting certain alternatives. Yet another possible application 
of the zero-probability assignment might be the case in which a decision 
maker, irrespective of their prior views, refuses to admit the real alternative. 
(For example, they have lied and then decide not to admit it.) Whether the 
behaviour of such pathological liars under noisy unravelling of information 
about the truth can be modelled using the zero-probability assignment 
approach here is an interesting open question.

\section{The role of noise}

In Brody \& Meier (2022) it is shown that even if a decision maker is unaware 
whether any given information is true or false, so long as they know the 
probability distribution of the fake news (represented by the time series 
$\{f_t\}$), then this is sufficient to eliminate the overall majority of the impact 
of fake news. In other words, anticipation of fake news is already a powerful 
antidote to its effects. While this feature is encouraging, it can also act against 
defending the truth, for, politicians nowadays often quote the phrase `fake 
news' to characterise inconvenient truth statements. Hence for those who 
believe in unfounded conspiracies, they anticipate truths being revealed 
which they perceive as false, and this anticipation also acts as a powerful 
antidote against accepting reality. Is there an alternative way of tackling the 
issue associated with strongly polarised clusters then? 

In this connection it is worth observing that the formation of domains and 
clusters described above is not uncommon 
in condensed matter physics of disordered systems. Here, atoms and 
molecules forming the matter interact with other atoms and molecules in 
their neighbourhoods. An atom, say, will then attempt to take the configuration 
that minimises the interaction energy with its neighbours (lower energy 
configurations are more stable in nature), but because this 
minimisation is a local operation, inconsistencies will emerge at large scales, 
and clusters 
of locally energy-minimising configurations will be formed. The boundary of 
different clusters, such as a domain wall, are called `defects' or `frustrations' 
in physics. 

To attempt to remove a defect, one can heat the system and then slowly 
cool it again. What the thermal energy does is to create a lot of noise, 
reconfiguring atoms and molecules 
in a random way so that after cooling back to the original state, the defect 
may be removed with a certain probability. This is essentially the idea of a 
Metropolis algorithm in a Monte Carlo simulation. Hence although noise is 
generally undesirable, it can play an important role in assisting a positive 
change, albeit only with a certain probability. 

There is an analogous situation that arises in biological processes 
(Trewavas 2012, 2016). Most 
biological processes are concerned with either processing the information 
about the environments, or else copying genetic information. 
In either case, noise 
is highly undesirable under normal circumstances. However, when a biological 
system is faced with an existential threat, then the situation is different. By 
definition, in such a circumstance, the conventional choices made by a 
biological system that would have been the correct ones under normal 
conditions are problematic, and it may be that for survival, the system 
must make a 
choice that \textit{a priori} seems incorrect. This is where noise can assist the 
system, to get over the threshold to reach unconventional choices. In other 
words, noise, as well as being a nuisance, is also what makes the system 
robust. 

Returning the discussion to disinformation, it should be evident that the main 
issue is not so much in the circulation of `fake news' \textit{per se}, but rather 
it is the coexistence of (a) polarised information clusters and (b) disinformation 
that creates real problems that are threats to democratic processes, or to public 
health. Hence to tackle the impact of disinformation a more effective way than 
the traditional `fact checking' strategy (which in itself of course has an essential 
role to play) seems to be in the dismantlement of the `defects' in the information 
universe, and this is where noise can potentially play an important role. Take, 
for example, the `odd one out' sample path shown on the left panel in 
Fig.~\ref{fig_2}. In spite of 
an extreme prior, this decision maker, assisted by unusually large values 
that the noise had taken, has been able to reduce the value of 
the prior associated with an incorrect decision over a relatively short time 
span. 

\begin{figure}[h]
      \centering
        \includegraphics[width=0.47\textwidth]{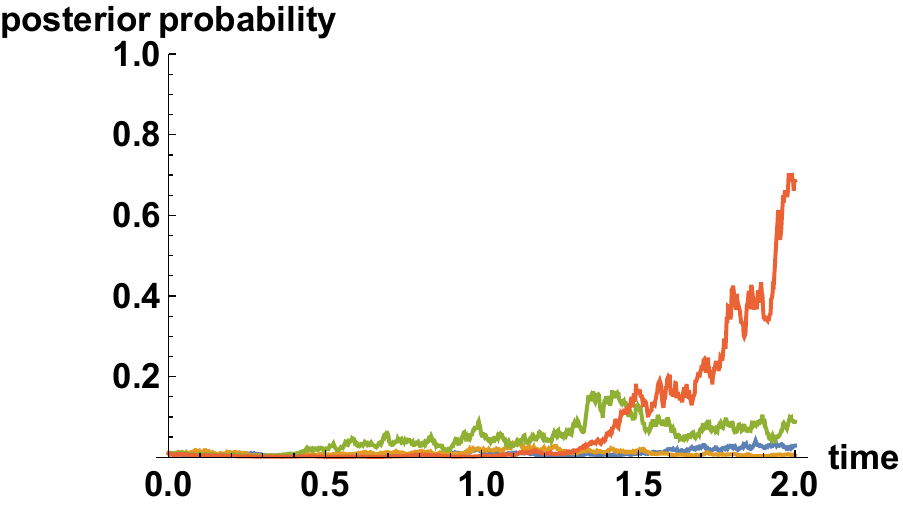} 
        \includegraphics[width=0.47\textwidth]{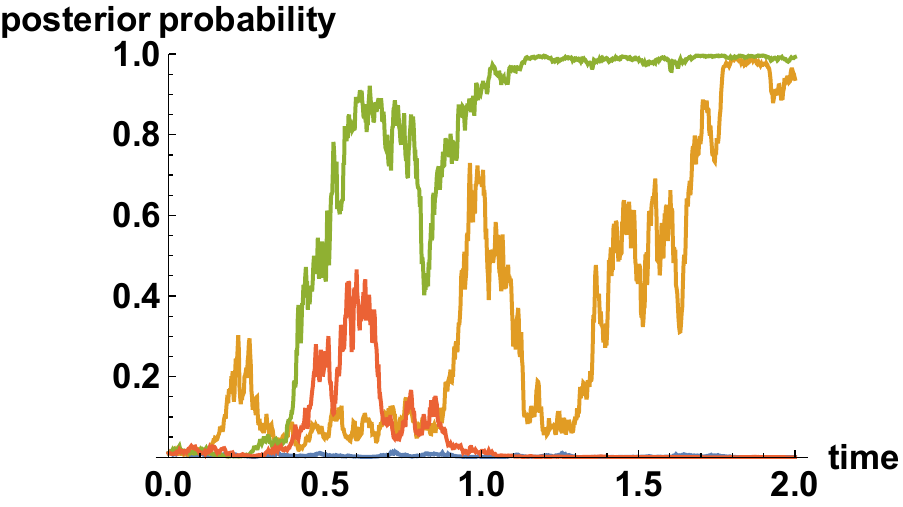}
        \caption{\textit{Tenacious Bayesian binary decision with enhanced noise}. 
        What happens to the tenacious Bayesian behaviour of the left panel in 
        Fig.~\ref{fig_2} if the noise level is enhanced in such a way that decision 
        makers are unaware of it? As in the example of Fig.~\ref{fig_2} , the 
        decision makers here have their priors set at 99\% for the choice $X=0$, 
        but the simulator has chosen $X=1$ to be the correct choice. 
        Plotted here are sample paths for the \textit{a posteriori} 
        probability that $X=1$. In the left 
        panel the noise level is doubled as compared to that of the left panel in 
        Fig.~\ref{fig_2}; whereas in the right panel it is quadrupled. In both cases, 
        the information flow rates are the same as that chosen in Fig.~\ref{fig_2}, 
        so no more reliable information is provided here for the Bayesian 
        decision makers. Nevertheless, the introduction of unknown noise 
        enhances the chance of arriving at the correct decision considerably 
        sooner, with positive probability. In particular, if the noise 
        level is quadrupled, then there is about 15\% chance that the noise will 
        assist such an escape from a false reality.}  
        \label{fig_3}
 \end{figure}

Of course, noise, having no bias, is unpredictable and the effect 
could have been the other way around. Nevertheless, without a substantial 
noise contribution the decision maker would have been stuck at a wrong 
place for a long time, and having a nonzero probability of an escape is 
clearly more desirable than no escape at all. In a similar vein, to dismantle 
an information cluster, rather than trying to throw factual information at it 
(which may not have an effect owing to the tenacious Bayesian phenomenon, 
and can also be costly), it may be more effective to significantly 
increase the noise level first, in such a way that decision makers are 
unaware of the increased level of noise, and then slowly removing it. The 
idea is to sufficiently confuse the misguided individuals, rather than forcing 
them to accept the facts from the outset. The result may be the resurgence 
of the original cluster, 
but there is a nonzero probability that the domain wall surrounding the 
cluster is dismantled. Putting it differently, an effective countermeasure 
against the negative impacts of disinformation might be the implementation 
of a real-life Metropolis algorithm or a simulated annealing (a slow cooling 
to reach a more stable configuration). 
 
As an example, in Fig.~\ref{fig_3} the impact of noise enhancement, when the 
decision makers are unaware of the noise manipulation, is shown, when the 
noise level is doubled and when it is quadrupled. If the orientation of noise 
goes against the truth, then the noise manipulation merely enforces the 
tenacious Bayeian effect more strongly 
(like one of the sample paths in blue on the right panel), but there is 
an equal probability that noise goes the other way, in which case the views 
of decision makers are altered considerably.

\section{Discussion: possible role of utility}

The theory of decision making under uncertainty is of course a 
well-established area of study in statistics (DeGroot 1970, Berger 1985). 
The theory outlined here departs from the traditional one by taking into 
consideration the flow of information that affects the perceptions of decision 
makers, thus allowing for an explicit dynamical characterisation of decision 
makings. This, in turn, opens up the possibility of engaging in a 
comprehensive case studies and scenario analysis. In this context, it also 
becomes evident how information manipulation in the form either of the 
dissemination of disinformation or noise adjustment can be built into the 
modelling framework, because the starting point of the analysis is the 
specification of the flow of information. 

Now in the context of statistical decision theory, the standard treatment 
presumes that an alternative is chosen if it maximises the expected utility, 
or perhaps if it minimises the expected loss (DeGroot 1970, Berger 1985). 
The utility function characterises the preference profile of a decision maker. 
While rational choice as characterised by maximising expected utility has 
been challenged (Kahneman \& Tversky 1979), the utility theory 
nevertheless works well in many applications. In particular, in the context of 
financial economics, correct valuations of assets are carried out by taking 
into account the impact of the utility. Putting the matter differently, when 
financial assets are priced by means of the expectation of the future return, 
this expectation is taken not with respect to the real-world probability, but 
rather, with respect to a risk-adjusted system of probabilities. 

It may be that analogously, when analysing, for instance, a voter's decision 
in an election 
it is more appropriate to consider a preference-adjusted probability associated 
with the utility profile of that voter, rather than the real-world probability. 
It is entirely possible, for instance, that some of the 
empirically observed phenomena such as confirmation bias can be 
explained even more accurately by combining the tenacious Bayesian 
behaviour with utility optimisation. Should 
this be the case, however, the information-based approach outlined here 
remains applicable; one merely has to reinterpret the probabilities slightly 
differently, but the formalism itself remains intact, and so are the conclusions. 

In summary, an information-based approach to characterising the dynamics 
of systems driven by information revelation has been elaborated here in 
some detail using simple decision-making scenarios, and the impact of 
information manipulation, including dissemination of disinformation, and how 
such concepts can be modelled in a scientifically meaningful manner, has 
been clarified. The effect of having an excessively high weight placed on a 
false belief -- called a tenacious Bayesian inference here -- is explained, and an 
extreme case of the effect, what one might call an alternative fact, is 
simulated to uncover their erratic 
characteristics. In particular, it is shown, based on the 
tenacious Bayesian behaviour, that confirmation bias can be explained, 
to an extent, within the Bayesian framework. 
Finally, a specific way of manipulating  
noise as a way of combatting the negative impact of disinformation is 
proposed. 

The information-based approach developed here not only 
allows for a systematic study of the behaviours of people under uncertain 
flow of information, but also can be implemented in practical applications. 
For sure some of the model parameters such as $\sigma$ and $f$ need 
not be controllable globally, especially in the context of a competition 
whereby one has no control over the strategies of the competitors. 
Nevertheless, there are means to estimate model parameters. For 
instance, in the context of an electoral competition, by studying the 
variability (volatility) of the opinion poll dynamics, the information flow rate 
$\sigma$ can be estimated quickly. Alternatively, model parameters may 
be calibrated from the response to changes in the strategy. For instance, 
in the context of marketing, one can ask how a 30\% increase in 
advertisement cost influenced on the sales figure. From such an analysis 
one can infer the level of information flow rate. At any rate, the mere fact 
that the information-based approach makes it possible to conduct a 
comprehensive impact studies and scenario analysis in itself is a huge 
advantage in developing informational strategies.

\vspace{0.45cm} 
\noindent 
{\bf Acknowledgements}.
The author thanks Lane Hughston, Andrea Macrina, 
David Meier and Bernhard Meister for discussion on related ideas.

\end{document}